\documentclass[reqno]{amsproc}

\usepackage{amssymb}

\usepackage{tikz}
\usetikzlibrary{patterns}
\usepackage{longtable}
\usetikzlibrary{chains}
\usepackage{float}

\newcommand{\wb}[1]{\overline{#1}}
\newcommand{\ub}[1]{\underline{#1}}

\newcommand{\nn}{\nonumber}

\newcommand{\be}{\begin{equation}}
\newcommand{\ee}{\end{equation}}
\newcommand{\bea}{\begin{eqnarray}}
\newcommand{\eea}{\end{eqnarray}}
\newcommand{\bse}{\begin{subequations}}
\newcommand{\ese}{\end{subequations}}
\newcommand{\benn}{\begin{equation*}}
\newcommand{\eenn}{\end{equation*}}
\newcommand{\beann}{\begin{eqnarray*}}
\newcommand{\eeann}{\end{eqnarray*}}

\newcommand{\al}{\alpha}
\newcommand{\dl}{\delta}

\newcommand{\ep}{\varepsilon}

\numberwithin{equation}{section}
\numberwithin{figure}{section}

\binoppenalty=\maxdimen
\relpenalty=\maxdimen

\begin{document}

\title[Singular dynamics of a $q$-difference Painlev\'{e} equation]{Singular dynamics of a $q$-difference Painlev\'{e} equation in its initial-value space}

\author{N. Joshi}
\address{School of Mathematics and Statistics, University of Sydney, NSW 2006, Australia}
\curraddr{}
\email{nalini.joshi@sydney.edu.au}
\thanks{}

\author{S.B. Lobb}
\address{School of Mathematics and Statistics, University of Sydney, NSW 2006, Australia}
\curraddr{}
\email{sarah.lobb@sydney.edu.au}
\thanks{}

\date{}

\begin{abstract}
We construct the initial-value space of a $q$-discrete first Painlev\'e equation explicitly and describe the behaviours of its solutions $w(n)$ in this space as $n\to\infty$, with particular attention paid to neighbourhoods of exceptional lines and irreducible components of the anti-canonical divisor. These results show that  trajectories starting in domains bounded away from the origin in initial value space are repelled away from such singular lines. However, the dynamical behaviours in neighbourhoods containing the origin are complicated by the merger of two simple base points at the origin in the limit. We show that these lead to a saddle-point-type behaviour in a punctured neighbourhood of the origin. \hfill\\
\begin{center}\textit{Dedicated to Rodney J. Baxter on the occasion of his 75th birthday.}\end{center}
\end{abstract} 

\maketitle

\section{Introduction}\label{section_introduction}
The discrete Painlev\'e equations are known to have a simple, beautiful geometric structure \cite{Sak2001}, but very little is known about their general solutions. Motivated by Boutroux's famous study \cite{Bou1913} of the first Painlev\'e equation $\text{P}_\text{I}$: $y_{tt}=6\,y^2-t$, as $|t|\to\infty$, we consider the limit $|\xi|\to\infty$ of its $q$-discrete version
\begin{align}\label{qP1}
q{\rm P}_{\rm I}: {w}\bigl(q\,\xi\bigr)\,w\bigl(\xi/q\bigr)&= \frac{w(\xi)-1/\xi }{w(\xi)^2},
\end{align}
where we assume $|q|>1$ for simplicity.  Such equations arise as reductions of Yang-Baxter maps \cite{ay:14,ptv:06} and multi-dimensionally consistent integrable lattice equations \cite{abs:03,jns:14}.

Starting at a given point in $\mathbb C$, repeated iteration of the equation on complex spirals (for $q\in{\mathbb C}$) leads to a sequence of solution values of a discrete Painlev\'e equation, but no information comparable to that found by Boutroux about global behaviours of solutions of $\text{P}_\text{I}$ is available in the plane of the independent variable $\xi\in\mathbb C$. Instead, we consider solution trajectories in the space of initial values, which is compactified by embedding in $\mathbb P^1\times \mathbb P^1$ and regularized by resolving all singularities\footnote{Note that $\mathbb P^1\times \mathbb P^1$ is the two-complex-dimensional space parametrized by pairs of homogeneous coordinates $([x_1:x_2],[y_1:y_2])$$=([\lambda\,x_1:\lambda\,x_2],[\mu\,y_1:\mu\,y_2])$ for all non-zero $(x_j, y_j)\in\mathbb C^2$, $j=1,2$, and non-zero $\lambda$, $\mu\in\mathbb C$.}. We denote the resulting initial value space by $\mathcal S$.

More precisely, writing $\xi=q^{n}\xi_{0}$, $w_{n}=w(\xi)$, $w_{n+1}=w(q\xi)$, $w_{n-1}=w(\xi/q)$, we study the solutions in ${\mathcal S}=\cup_{n_0\in\mathbb C} S_{n_0}$, where the fibre $S_{n_0}$ is locally described by affine coordinates $(w_{n_0},w_{n_0-1})$ and resolved by blowing up 8 base points. Equation \eqref{qP1} is symmetric under $w_{n+1}\leftrightarrow w_{n-1}$. 
In the remainder of the paper, we study the equivalent forward and backward systems
\begin{equation}\label{qp1}
\begin{pmatrix} 
	\wb{u} \\
	\wb{v}
	\end{pmatrix}
 =  \begin{pmatrix} 
		\displaystyle\frac{q\wb{v}-t}{q u\wb{v}^2}\\ 
  		\displaystyle\frac{u-t}{u^2v}
		\end{pmatrix},
\qquad
\begin{pmatrix} 
	\ub{u} \\
	\ub{v}
	\end{pmatrix}	
	 =  \begin{pmatrix} 
			\displaystyle\frac{v-qt}{uv^2}\\ 
  			\displaystyle\frac{\ub{u}-q^2t}{\ub{u}^2v} 
		\end{pmatrix}
\end{equation}
where $t:=1/\xi$, $u_{n}:=w_{2n}$, $v_{n}:=w_{2n-1}$. (Note that bar now denotes $t\mapsto t/q^2$.)

To our knowledge, the term ``discrete Painlev\'e equation'' was proposed by Fokas, Its and Kitaev \cite{FokItsKit1991} for another discrete version of $\text{P}_\text{I}$, which arises from transformations of the classical fourth Painlev\'e equation. The latter equation is an additive-type discrete equation, where coefficient functions are linear in $n$ instead of exponential, i.e., $\xi=\xi_0\,q^n$, as in Equation \eqref{qP1}. 

Many discrete Painlev\'e equations have now been identified. Following Okamoto's construction for  the classical Painlev\'e equations \cite{Oka79}, Sakai \cite{Sak2001} showed how to construct initial value spaces of discrete Painlev\'e equations as rational surfaces obtained by 9-point resolution of $\mathbb P^2$ or equivalently, 8-point resolution of $\mathbb P^1\times \mathbb P^1$. In particular, Sakai showed that the initial value space associated with Equation \eqref{qp1} is described by the affine Weyl group $A_{7}^{(1)}$. 

Qualitative information about solutions of \eqref{qp1} is valuable because Nishioka \cite{Nis2010} showed that they are higher transcendental functions, which cannot be expressed in terms of basic or $q$-special functions or in terms of solutions of first-order or linear second-order difference equations.  We describe solutions by studying the limit $|\xi|\to\infty$, or equivalently $|t|\to0$ on its initial value surface.

\subsection{Autonomous Limit}\label{auto} In the limit $|t|\to0$, the system \eqref{qp1} becomes 
\begin{equation}\label{qp1 auto}
\begin{pmatrix} 
	\wb{u} \\
	\wb{v}
	\end{pmatrix}
 =  \begin{pmatrix} 
		v\\ 
  		\displaystyle\frac{1}{u v}
		\end{pmatrix},
\qquad
\begin{pmatrix} 
	\ub{u} \\
	\ub{v}
	\end{pmatrix}	
	 =  \begin{pmatrix} 
			\displaystyle\frac{1}{u v}\\ 
  			u
		\end{pmatrix}
\end{equation}
or, equivalently 
$ \overline{w}\,{w}\,\underline{w} =1$, for either $w=u$ or $v$. This map is one of the periodic cases of the QRT mappings studied by Tsuda \cite{Tsu2004}, which is periodic with period 3, for any initial value. The autonomous system has an invariant given by
\begin{equation}\label{eq: ham auto}
K(u, v) =\frac{u^2v^2+u+v}{uv}
\end{equation}
i.e., $K(\overline{u}, \overline v)-K(u, v)=0=K(u, v)-K(\underline u, \underline v)$ when $u$ and $v$ satisfy Equations \eqref{qp1 auto}. The singularities of the invariant curve (i.e., points where the gradient of $K(u, v)$ vanishes) are given by $(u, v)=(\omega, \omega)$ where $\omega^3=1$, and $(u, v)=(0, \infty)$, $(u,v)=(\infty, 0)$. These lie on the invariant curve only if $K=3\omega^2$. Therefore, there are 3 singular fibres in the pencil of curves \eqref{eq: ham auto}. 

It is interesting to note that the so called \emph{tronqu\'ee} solutions identified by Boutroux for the first Painlev\'e equation also approach double zeroes of the latter's autonomous invariant. In this sense, the solutions that approach the equilibrium values $(u, v)=(\omega, \omega)$ of the autonomous $q\text{P}_\text{I}$ act as discrete analogues of the tronqu\'ee solutions.
In this paper, however, we focus on the generic solutions of the dynamical system \eqref{qp1}.

\subsection{Fixed points} Fixed points are important for studying limiting solutions. The cases where the solutions of Equation \eqref{qP1} approach a steady state as $n\to+\infty$ have been considered in \cite{Jos2014,Ohy2010}. Joshi \cite{Jos2014} showed that there exists a true solution satisfying $|w|\to0$ as $n\to+\infty$, which is asymptotic to a divergent asymptotic series expansion in powers of $1/\xi$, but is not a singularity of the limiting invariant $K$. This vanishing, unstable solution was called the \textit{quicksilver} solution in  \cite{Jos2014} and its trajectory lies in a neighbourhood of the origin in $\mathcal S$, which is punctured at two base points (called $b_1$, $b_2$ below) that approach the origin. (See Figure \ref{fig_sing_dom} for a typical such neighbourhood.) Our results show that the dynamics of Equations \eqref{qp1} in this punctured neighbourhood resemble that near a saddle point in the initial value space $\mathcal S$.
   
\subsection{Outline of the Paper} The structure of the paper is as follows. In Section \ref{section_res_sing} we construct the initial-value space $\mathcal S$ and provide explicit coordinate charts to describe it completely; key points are illustrated here, while full details are given in Appendix \ref{appendix_res_sing}. We also describe the irreducible components of its anti-canonical divisor, with explicit details given in Appendix \ref{appendix_divisors}. In Section \ref{section_soln1278} we examine the action of the mapping on exceptional lines, with details given in Appendix \ref{appendix_elines}. We then describe approximate solutions near exceptional lines, with particular attention paid to a neighbourhood of the origin in Section \ref{section:e1}. A conclusion and summary are provided in Section \ref{section_conclusion}.

\section{The Initial Value Space}\label{section_res_sing}
In this section, we describe the major features of the initial value space $\mathcal S$, obtained by resolution of singularities of the dynamical system \eqref{qp1}. The details of the calculations are provided in Appendices  \ref{appendix_res_sing} and \ref{appendix_divisors}.

\subsection{Resolution of the Initial Value Space}
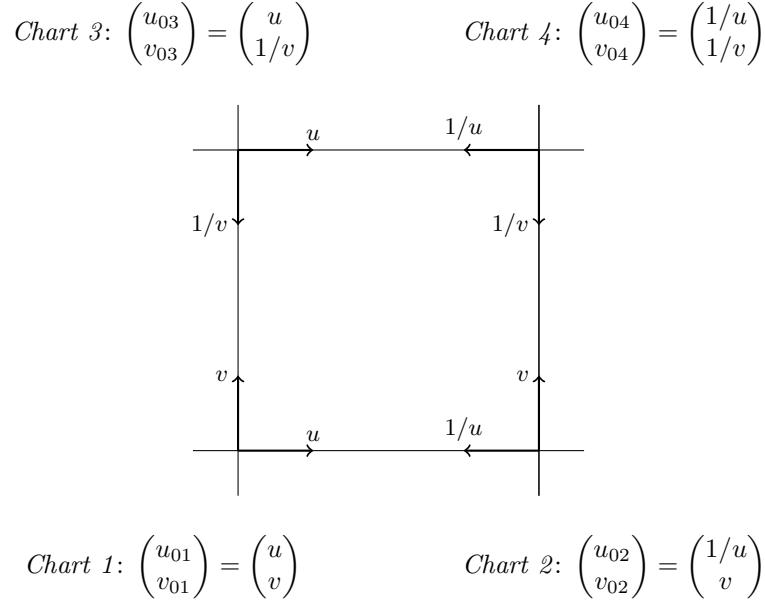
\begin{figure}
\begin{center}
\begin{tikzpicture}[scale=2]
    \draw (0,2.3) -- (0,-0.3) ;
    \draw (-0.5,-0.5) node [anchor=north] {{\em Chart 1}\/: $\begin{pmatrix} 
  u_{01}\\
  v_{01}
	\end{pmatrix}	
  =\begin{pmatrix} 
	u\\
	v
  \end{pmatrix}$};
    \draw[->,thick] (0,0) -- (0,0.5) node[anchor=east] {\small $v$};
    \draw (2.3,0) -- (-0.3,0) ;
    \draw[->,thick] (0,0) -- (0.5,0) node[anchor=south] {\small $u$};
	
    \draw[->,thick] (0,2) -- (0,1.5) node[anchor=east] {\small ${1}/{v}$};
    \draw (2.3,2) -- (-0.3,2) ;
    \draw[->,thick] (0,2) -- (0.5,2) node[anchor=south] {\small $u$};
    \draw (-0.5,2.5) node [anchor=south] {{\em Chart 3}\/: $\begin{pmatrix} 
  	u_{03}\\
  	v_{03}
	\end{pmatrix}
	= \begin{pmatrix} u\\ 
	1/v
	\end{pmatrix}$};

    \draw (2,2.3) -- (2,-0.3);
    \draw[->,thick] (2,2) -- (2,1.5) node[anchor=east] {\small ${1}/{v}$};
    \draw (2,2.3) -- (2,-0.3) ;
    \draw[->,thick] (2,2) -- (1.5,2) node[anchor=south] {\small $1/u$};
    \draw (2.5,2.5) node [anchor=south] {{\em Chart 4}\/: $\begin{pmatrix} 
  	u_{04}\\
	v_{04}
	\end{pmatrix}	
  =\begin{pmatrix} 
  	{1}/{u}\\ 
		{1}/{v}
      \end{pmatrix}$};

    \draw[->,thick] (2,0) -- (2,0.5) node[anchor=east] {\small $v$};
    \draw[->,thick] (2,0) -- (1.5,0) node[anchor=south] {\small $1/u$};
    \draw (2.5,-0.5) node [anchor=north] {{\em Chart 2}\/: $\begin{pmatrix} 
     u_{02}\\
     v_{02}
	\end{pmatrix}	
  =\begin{pmatrix} 
	{1}/{u}\\ 
	v
  \end{pmatrix}$};

\end{tikzpicture}
\caption{The four coordinate charts of $\mathbb{P}^1\times \mathbb{P}^1$}\label{regions}
\end{center}
\end{figure}

Consider the dynamical system \eqref{qp1} in $\mathbb P^1\times \mathbb P^1$, which is covered by four charts described in Figure \ref{regions}. Singularities occur where the numerator and denominator vanish simultaneously in the definition of at least one of $\overline u$, $\overline v$, $\underline u$, or $\underline v$. We call these \textit{base points}, following \cite{Dui2010}. We resolve these in the standard way\/\footnote{We recall here that the standard operation of resolution or blowing up a base point at $(\alpha,\beta)$ means replacing $(u, v)$ by two new coordinate charts $\tilde u=(u-\alpha)/(v-\beta), \tilde v= v-\beta$, or $\hat u=(u-\alpha), \hat v=(v-\beta)/(u-\alpha)$. This has the effect of replacing the base point by an {\em exceptional} line. Each resolution lowers the self-intersection number of the line containing the base point  by unity. In the chart $(\tilde u, \tilde v)$, this line is defined by $\tilde v=0$ and parametrized by $\tilde u$, whilst in chart $(\hat u, \hat v)$, the line is defined by $\hat u=0$ and parameterized by $\hat v$.  For this and other standard constructions of algebraic geometry, the reader is referred to \cite{Har1977}.} to obtain successive charts $(u_{ij},v_{ij})$, where $i$ corresponds to the number of blow-ups and $j$ corresponds to a chart indicated in Figure \ref{regions}. 

The definition of $\overline v$ and $\underline u$ in Equations \eqref{qp1} show that there are two base points in Chart 1:
\be
 b_{1}: \left(\begin{array}{c} u_{01} \\ v_{01} \end{array}\right) = \left(\begin{array}{c} t \\ 0 \end{array}\right),
 \qquad  
 b_{2}: \left(\begin{array}{c} u_{01} \\ v_{01} \end{array}\right) = \left(\begin{array}{c} 0 \\ q t \end{array}\right),
\ee
while there is one base point in each of Charts 2 and 3 respectively:
\be
 b_{3}: \left(\begin{array}{c} u_{02} \\ v_{02} \end{array}\right)  = \left(\begin{array}{c} 0 \\ 0 \end{array}\right),\qquad
  b_{4}: \left(\begin{array}{c} u_{03} \\ v_{03} \end{array}\right) = \left(\begin{array}{c} 0 \\ 0 \end{array}\right).
\ee
The explicit resolutions of these base points are provided in Appendix \ref{appendix_res_sing}. 

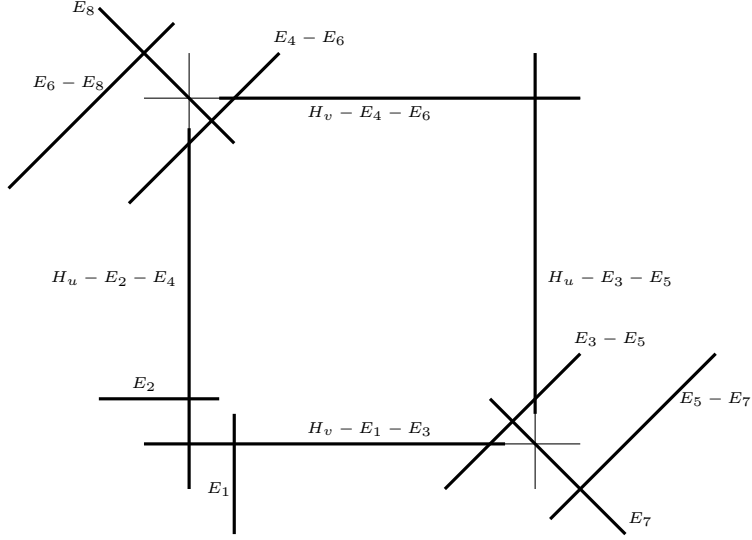
\begin{figure}
\begin{center}
\begin{tikzpicture}[scale=2]
    \draw (2.3,-0.3) -- (2.3,2.6) ;
    \draw (0,-0.3) -- (0,2.6) ;
    \draw (-0.3,0) -- (2.6,0) ;
    \draw (-0.3,2.3) -- (2.6,2.3);
   \draw [very thick] (1.7, -0.3) -- (2.6,0.6);
   \draw [very thick] (-0.4, 1.6) -- (0.6,2.6);
    \draw [very thick] (-0.6, 2.9) -- (0.3,2);
    \draw [very thick] (2.4, -0.5) -- (3.5, 0.6);
    \draw [very thick] (2, 0.3) -- (2.9,-0.6);
    \draw [very thick] (-1.2, 1.7) -- (-0.1, 2.8);
    \draw [very thick] (0,-0.3) -- (0,2.1) ;
    \draw [very thick] (0.2,2.3) -- (2.6,2.3);
    \draw [very thick] (2.3,0.2) -- (2.3,2.6) ;
    \draw [very thick] (-0.3,0) -- (2.1,0) ;

    \draw [very thick] (0.3,-0.6) -- (0.3,0.2) ;
    \draw [very thick] (-0.6,0.3) -- (0.2,0.3) ;
     \draw (-0.5,1.1) node {\tiny $H_{u}-E_{2}-E_{4}$};
     \draw (2.8,1.1)  node {\tiny $H_{u}-E_{3}-E_{5}$};
     \draw (1.2,2.2) node {\tiny $H_{v}-E_{4}-E_{6}$};
     \draw (1.2, 0.1) node {\tiny $H_{v}-E_{1}-E_{3}$};
     \draw (0.2,-0.3) node  {\tiny $E_{1}$};
    	\draw (-0.3,0.4) node {\tiny $E_{2}$};
    	\draw (2.8,0.7) node {\tiny $E_{3}-E_5$};
    	\draw (3.5, 0.3) node {\tiny $E_{5}-E_7$};
    	\draw (3, -0.5) node {\tiny $E_7$};
	
    	\draw (0.8,2.7) node {\tiny $E_{4}-E_6$};
    	\draw (-0.8, 2.4) node {\tiny $E_{6}-E_8$};
    	\draw (-0.7, 2.9) node {\tiny $E_8$};

 \end{tikzpicture}
\caption{The rational surface $\mathcal S$ obtained from the resolution of singularities in Equation \eqref{qp1} indicating proper transforms of relevant curves.}
\label{figEl8}
\end{center}
\end{figure}

The resolution of $b_1$ and $b_2$ lead to no further singularities (see \S \ref{a:b1}-\ref{a:b2}). However, the resolutions of $b_3$ and $b_4$ lead to four further base points, which are given in terms of coordinate charts defined in \S \ref{a:b3}-\ref{a:b4} by 
\begin{align}
 &b_{5}: \left(\begin{array}{c} u_{31} \\ v_{31} \end{array}\right) = \left(\begin{array}{c} 0 \\ 0 \end{array}\right),\ \ 
 b_{6}: \left(\begin{array}{c} u_{42} \\ v_{42} \end{array}\right) = \left(\begin{array}{c} 0 \\ 0 \end{array}\right),\\
 &b_{7}: \left(\begin{array}{c} u_{51} \\ v_{51} \end{array}\right) = \left(\begin{array}{c} -q \\ 0 \end{array}\right),  
 b_{8}: \left(\begin{array}{c} u_{61} \\ v_{61} \end{array}\right) = \left(\begin{array}{c} -q \\ 0 \end{array}\right).
\end{align}
The resolution of $b_3$ leads to $b_5$, $b_4$ leads to $b_6$, while those of $b_5$ and $b_6$ lead to $b_7$ and $b_8$ respectively.
The details can be found in \S \ref{a:b3}-\ref{a:b8}.  The exceptional lines $E_j$ replacing each base point $b_j$, $j=1, \ldots, 8$ are drawn schematically in Figure  \ref{figEl8}, where $H_{u}$ denotes a line where $u$ is constant and $H_{v}$ denotes a line where $v$ is constant.

\subsection{Divisors}\label{section_divisors}
The equivalence classes of lines $H_{u},H_{v},$ and $E_{i}$ for $i=1,..,8$ form the standard basis of the Picard group Pic($\mathcal S$) \cite{Har1977}. This is equipped with a symmetric bilinear form $(\cdot,\cdot)$ on elements of Pic($\mathcal S$) given by 
\begin{enumerate}
 \item $(H_{u}, H_{u}) = (H_{v}, H_{v}) = (H_{u}, E_{i}) = (H_{v}, E_{i}) = 0$,
 \item $(H_{u}, H_{v}) = 1$,
 \item $(E_{i}, E_{j}) = -\dl_{ij}$.
\end{enumerate}

From the resolution described above (see Figure \ref{figEl8}), it follows that the nodal curves with self-intersection -2 are
\begin{align*}
 D_{1} &= H_{v}-E_{1}-E_{3}, &D_{5} &= H_{v}-E_{4}-E_{6},\\
 D_{2} &= E_{3}-E_{5}, &D_{6} &= E_{6}-E_{8},\\
 D_{3} &= E_{5}-E_{7},&D_{7} &= E_{4}-E_{6},\\
 D_{4} &= H_{u}-E_{3}-E_{5},&D_{8} &= H_{u}-E_{2}-E_{4}.
 \end{align*}
These form irreducible components of the anti-canonical divisor, defined by \hfill \\$-K_{\mathcal S}$$=2H_{u}$$+2H_{v}$$-\sum_{i=1}^{8}{E_{i}}$. By the definitions above, we have $-K_{\mathcal S}=\sum_{j=1}^8 D_j$. 

\begin{figure}
\begin{center}
\begin{tikzpicture}[start chain=circle placed {at=(\tikzchaincount*45:2)}]
  \foreach \i in {1,...,8}
    \node [draw,circle,on chain,join] {\tiny $D_\i$};
    \draw (circle-8) -- (circle-1);
\end{tikzpicture}
\caption{Intersection diagram of $\{D_j\}_{j=1}^8$, which is equivalent to the Dynkin diagram of $A_7^{(1)}$.}
\label{figDynkin}
 \end{center}
 \end{figure}
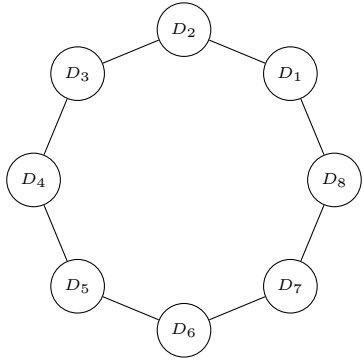

From the intersection rules on the Picard group given above, it follows that $(D_1,D_2)=1=(D_1,D_8)$, while $(D_1, D_j)=0$ for $j=3, 4, 5, 6, 7$. Further details about the intersections of $D_j$ and the corresponding generalized Cartan matrix are provided in Appendix \ref{appendix_divisors}. By representing each $D_j$ as a node and connecting two nodes $D_i$, $D_j$ by a simple edge if $(D_i, D_j)\not=0$, we obtain the Dynkin diagram given in Figure \ref{figDynkin}, which represents the affine Weyl group $A_7^{(1)}$.
By considering its orthogonal complement in Pic($\mathcal S$), we also show explicitly in Appendix \ref{appendix_divisors} that Equation \eqref{qp1} has a symmetry group given by $A_1^{(1)}$.
\section{Dynamics}

Let the forward iteration given in Equation \eqref{qp1} on $\mathcal S$ be denoted by $\varphi$. We study iterations of points near the components $D_{j}$ of the anti-canonical divisor in \S \ref{s:dynamics_div}, near the exceptional lines $E_{1}$, $E_{2}$, $E_{7}$, $E_{8}$ in \S \ref{section_soln1278} and in a  region around the origin punctured at $E_1$ and $E_2$ in \S \ref{section:e1}. The section ends with the explicit linearisation of the map on the Picard lattice.

\subsection{Dynamics near the irreducible components $D_{i}$}\label{s:dynamics_div}
\begin{figure}[H]
\begin{center}
\begin{tikzpicture}[start chain=circle placed {at=(\tikzchaincount*90:2)},every join/.style=thick,->]
 \node [on chain, join] {\tiny $D_{1}$};
 \node [on chain, join] {\tiny $D_{7}$};
 \node [on chain, join] {\tiny $D_{5}$};
 \node [on chain, join] {\tiny $D_{3}$};
    \draw[thick,->] (circle-4) -- (circle-1);
\end{tikzpicture}
\hspace{1cm}
\begin{tikzpicture}[start chain=circle placed {at=(\tikzchaincount*90:2)},every join/.style=thick,->]
 \node [on chain, join] {\tiny $D_{4}$};
 \node [on chain, join] {\tiny $D_{2}$};
 \node [on chain, join] {\tiny $D_{8}$};
 \node [on chain, join] {\tiny $D_{6}$};
    \draw[thick,->] (circle-4) -- (circle-1);
\end{tikzpicture}
\caption{Action of $\varphi$ on divisors.}
\label{figAction_div}
 \end{center}
 \end{figure}
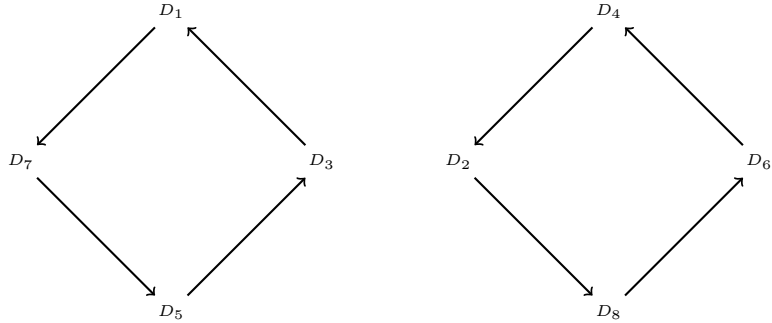

The iteration of these components under $\varphi$ is computed in Appendix \ref{appendix_divisors}. For example, consider $D_1=H_u-E_1-E_3$. Given $t=t_0$, a finite initial point near $D_1$ lies near the line $v_{01}=0$ (with $u_{01}\not=t_0$), so that $v_{01}(t_{0})=\dl v'_{01}(t_{0})+O(\dl^2)$ for some finite $v'_{01}(t_{0})$, and small $0<\dl\ll 1$.  The results of  Appendix \ref{appendix_divisors} show that
\be
	\left\{\begin{array}{rcl} u_{01}(t_{0}/q^8) & \sim & -\dfrac{u_{01}(t_{0})\,t_0^3}{q^{14}},\\ 
					   v_{01}(t_{0}/q^8) & \sim & \dl\dfrac{q v'_{01}(t_{0})t_{0}^2\left(q^6+u_{01}(t_{0})t_{0}^2\right)}{(u_{01}(t_{0})-t_{0})}, \end{array} \right. 
\ee
which again lies near $D_1$ after iterating forward 4 steps. Arguments similar to these in Appendix \ref{appendix_divisors} show that the set $D_{1},\dots,D_{8}$ is closed under the action of $\varphi$, the time evolution of $D_j$ is periodic and, moreover, their iteration occurs in two orbits of period four each, as shown in Figure \ref{figAction_div}.

\subsection{Dynamics near exceptional lines $E_{1}$, $E_{2}$, $E_{7}$, $E_{8}$}\label{section_soln1278}
As shown in Appendix \ref{appendix_elines}, we have the following behaviour under forward iteration
\be\label{chain1}
 E_{7} \to E_{1} \to H_{u}+2H_{v}-E_{2}-E_{3}-E_{4}-E_{5}-E_{7}, \quad\textrm{and}\quad E_{2} \to E_{8} \to H_{v}-E_{2},
\ee
and under backward iteration we have
\be\label{chain2}
 H_{u}-E_{1} \leftarrow E_{7} \leftarrow E_{1}, \quad\textrm{and}\quad 2H_{u}+H_{v}-E_{1}-E_{3}-E_{4}-E_{6}-E_{8}\leftarrow E_{2} \leftarrow E_{8} .
\ee
This means that under forward and backward iterations of these exceptional lines, we are sent eventually to regular points. There are certain special cases: for example $(u_{11},v_{11})=(0,0)$ on $E_{1}$ is mapped under forward iteration to the point $(u_{71},v_{71})$$=(-q^2\xi,0)$ on $E_{7}$. This point is mapped through the chain \eqref{chain1} once more, before emerging into regular space.

\subsection{Dynamics near the origin}\label{section:e1}
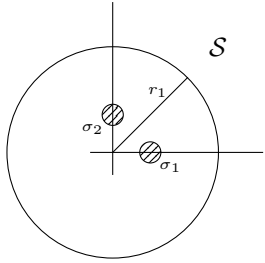
\begin{figure}[H]
\begin{center}
\begin{tikzpicture}
    \draw (0,2) -- (0,-0.3) ;
    \draw (2, 0) -- (-0.3,0) ;
    \draw (0,0) circle [radius=40pt];
    \draw[pattern=north east lines] (0.5,0) circle [radius=4pt] node [anchor=north west] {\tiny $\sigma_1$};
    \draw[pattern=north east lines] (0,0.5) circle [radius=4pt] node [anchor=north east] {\tiny $\sigma_2$};
    \draw (0,0) -- (1,1) ;
    \draw (0.6,0.8) node {\tiny $r_1$};
    \draw (1.4,1.4) node {$\mathcal S$};
    \end{tikzpicture}
\caption{Punctured domain $U$}
\label{fig_sing_dom}
 \end{center}
 \end{figure}

Here we consider dynamics in a neighbourhood of the origin in initial value space. Assume $d_{r_1}(0,0)$ is a disk of radius $r_1$ around the origin in $\mathcal S$ for sufficiently small $r_1>0$.  We assume that $t$ lies in a small domain $d_{r_2}(t_0)$, around a given point $t_0\ll 1$, for sufficiently small $r_2>0$ and work in the domain $U\subset d_{r_1}(0,0)$ lying in $\mathcal S$, which is  punctured at $(t, 0)$ and $(0,q t)$ . We denote the small punctured domain around $(t, 0)$ by $\sigma_1$ and that around $(0, q t)$ by $\sigma_2$. An example of such a domain is drawn in Figure \ref{fig_sing_dom}. 
The mapping of $\sigma_j$, $j=1,2$ under $\varphi$ are obtained by the results in Appendix \ref{appendix_elines}. These show that the exceptional lines in each region act as repellers, and that iterates move away from each region. 

In the domain $U$, for sufficiently small $(u, v)$, we find
\begin{align}\label{qp1_0}
	&\wb{u} \sim -\,\frac{u\,v}{t}&&\ub{u}\sim -\,\displaystyle\frac{qt}{uv^2}\\
	&\wb{v} \sim -\, \frac{t}{u^2v}&& \ub{v}\sim -\,\displaystyle\frac{u\,v}{q\,t}
\end{align}
These show that $U$ is mapped to a domain with large values of $v$ (i.e., towards the origin in chart 3, which corresponds to $(u, v)=(0, \infty)$) by the forward map, while it is mapped to a domain with large values of $u$ (i.e., towards the origin in chart 2, which corresponds to $(u, v)=(\infty, 0)$) by the backward map. In other words, under the forward map, a domain with large values of $v$ and finite $u$ is attracted to the origin, while values of $u$ and $v$ lying in $U$ are repelled away to a domain with large values of $u$ and finite $v$. That is, the origin is a saddle-point of the dynamical system. This result is consistent with the origin being a repelling fixed point of the autonomous map, and with the quicksilver solution being an unstable solution, as shown in \cite{Jos2014}. 

\subsection{Summary of behaviours}
Appendix \ref{appendix_divisors} contains explicit calculations of the mappings of the irreducible components $D_j$ of the anti-canonical divisor, and Appendix \ref{appendix_elines} contains explicit calculations of the mappings of the remaining exceptional lines. 

Based on these results, we can express the action of $\varphi$ on elements of the standard basis of the Picard group in matrix form:
\be
 \varphi\left( \begin{array}{c} H_{u}\\ H_{v}\\ E_{1}\\ E_{2}\\ E_{3}\\ E_{4}\\ E_{5}\\ E_{6}\\ E_{7}\\ E_{8} \end{array} \right)
  = \left( \begin{array}{rrrrrrrrrr}   1 & 2 & 0 & -1 & -1 & -1 & -1 & 0 & 0 & 0\\
  						2 & 3 & 0 & -2 & -2 & -1 & -1 & -1 & -1 & 0\\
						1 & 2 & 0 & -1 & -1 & -1 & -1 & 0 & -1 & 0\\
						0 & 0 & 0 & 0 & 0 & 0 & 0 & 0 & 0 & 1\\
						1 & 1 & 0 & -1 & -1 & -1 & 0 & 0 & 0 & 0\\
						1 & 2 & 0 & -1 & -1 & -1 & -1 & -1 & 0 & 0\\
						0 & 1 & 0 & 0 & -1 & 0 & 0 & 0 & 0 & 0\\
						1 & 1 & 0 & -1 & -1 & 0 & -1 & 0 & 0 & 0\\
						0 & 0 & 1 & 0 & 0 & 0 & 0 & 0 & 0 & 0\\
						0 & 1 & 0 & -1 & 0 & 0 & 0 & 0 & 0 & 0 \end{array} \right)
      \left( \begin{array}{c} H_{u}\\ H_{v}\\ E_{1}\\ E_{2}\\ E_{3}\\ E_{4}\\ E_{5}\\ E_{6}\\ E_{7}\\ E_{8} \end{array} \right)
\ee
where $\varphi:\textrm{Pic}(\mathcal S)\to\textrm{Pic}(\mathcal S)$ is the action of forward iteration in $n$. Because all the eigenvalues of this matrix are  roots of unity, its \emph{algebraic entropy} (see Bellon and Viallet \cite{BelVia1999}) vanishes. This shows that the complexity of the dynamics of this system does not grow exponentially in the time step $n$.

\section{Conclusions}\label{section_conclusion}
In this paper, we explicitly constructed the initial value space of the $q$-discrete first Painlev\'e system \eqref{qp1} as $t\to0$, and for that of the scalar equation \eqref{qP1}, in the limit $\xi\to\infty$. We  provided the coordinate charts of the space of initial values in detail, identifying exceptional lines and irreducible components of the anti-canonical divisor and, based on these, showed how to deduce qualitative information about the dynamics in the limit. 

We found that the union of exceptional lines is a repeller for the flow; i.e., if a solution is near an exceptional line at any given time, after one time step it is immediately mapped to a different region. This is analogous to the results for Painlev\'e I found in \cite{DuiJos2011}. The set of the irreducible components of the anti-canonical divisor is invariant under the time flow and the dynamical system is periodic on the level of each such component. 

Nevertheless, the solutions are neither simple or periodic. In particular, we showed that the origin acts as a saddle point for the generic flow. The remaining fixed point solutions, which approach the singularities on three fibres of the autonomous pencil \eqref{eq: ham auto} (see \S \ref{auto}) are the discrete analogues of the so called \emph{tronqu\'ee} solutions identified by Boutroux for the first Painlev\'e equation, which also approach double points of the latter's autonomous Hamiltonian.   

The dynamics of most other $q$-discrete Painlev\'e equations in limits when $t\to0$ or $t\to\infty$ remain unknown. In particular, it would be useful to study the $q$-discrete Painlev\'e equations known as $q{\rm P}_{\rm II}$ and $q{\rm P}_{\rm III}$, which have $q{\rm P}_{\rm I}$ as a degenerate limit. 

Finally, since these $q$-discrete Painlev\'e equations arise as reductions of Yang-Baxter maps, it would be interesting to relate their limiting behaviours to the physical setting in which such maps play a role.

\appendix

\section{Blow up of base points}\label{appendix_res_sing} 

Here we provide explicit details of the process of resolution of each of the 8 base points of $q{\rm P}_{\rm I}$.


\subsection{Blow up of base point $b_{1}$}\label{a:b1}

This base point arises in Chart 1 when we iterate the map forward.

Define new coordinates
\be
 \left(\begin{array}{c} u_{11}\\ v_{11} \end{array}\right) = \left(\begin{array}{c} (u_{01}-t)/v_{01} \\ v_{01} \end{array}\right) \quad \Rightarrow \quad
 \left(\begin{array}{c} u_{01}\\ v_{01} \end{array}\right) = \left(\begin{array}{c} u_{11}v_{11}+t \\ v_{11} \end{array}\right).
\ee

In these coordinates, Equations \eqref{qp1} become
\bse\label{eqn11}
\bea
\left\{\begin{array}{rcl} \wb{u} & = &\dfrac{\left(u_{11} v_{11}+t\right) \left(q u_{11} - u_{11}^2 v_{11}^2t - 2u_{11} v_{11}t^2 - t^3\right)}{q u_{11}^2}\\ 
					   \wb{v} & = &\dfrac{ u_{11}}{\left(u_{11} v_{11}+t\right){}^2} \end{array} \right. 
\eea
and
\bea
					            \left\{\begin{array}{rcl} \ub{u} & = & \dfrac{v_{11}-qt}{v_{11}^2 \left(u_{11} v_{11}+t\right)}\\ 
					            &&\\
         					   \ub{v} & = & \dfrac{v_{11} \left(u_{11} v_{11}+t\right) \left(v_{11} - q^2 u_{11} v_{11}^3t -qt - q^2 v_{11}^2t^2\right)}{ \left(v_{11}-qt\right){}^2} \end{array}\right.
 \eea
\ese
where the base point $b_{1}$ is replaced by the exceptional line $E_{1}$ defined by $v_{11}=0$. There are no new base points in this chart.

To look in the other chart, define new coordinates
\be
 \left(\begin{array}{c} u_{12}\\ v_{12} \end{array}\right) = \left(\begin{array}{c} u_{01}-t \\ v_{01}/(u_{01}-t) \end{array}\right)\quad \Rightarrow \quad
 \left(\begin{array}{c} u_{01}\\ v_{01} \end{array}\right) = \left(\begin{array}{c} u_{12}+t \\ u_{12}v_{12} \end{array}\right).
\ee
The equations in these coordinates become
\bse\label{eqn12}
\bea
\left\{\begin{array}{rcl} \wb{u} & = &\dfrac{\left(u_{12}+t\right) v_{12} \left(q - u_{12}^2 v_{12}t - 2u_{12}v_{12}t^2 - v_{12}t^3\right)}{q}\\
					   \wb{v} & = &\dfrac{1}{\left(u_{12}+t\right){}^2 v_{12}}\end{array} \right.
	\eea
	and
	\bea				   
					            \left\{\begin{array}{rcl} \ub{u} & = & \dfrac{u_{12} v_{12}-qt}{u_{12}^2 \left(u_{12}+t\right) v_{12}^2}\\ 
         					  \ub{v} & = & \dfrac{u_{12} \left(u_{12}+t\right) v_{12} \left(u_{12} v_{12} - q^2 u_{12}^3 v_{12}^2t -qt - q^2 u_{12}^2 v_{12}^2t^2\right)}{\left(u_{12} v_{12}-qt\right){}^2} \end{array}\right.\nn
\eea
\ese
where the base point $b_{1}$ is replaced by the exceptional line $E_1$ defined by $u_{12}=0$. There are no new base points in this chart.


\subsection{Blow up of base point $b_{2}$}\label{a:b2}
Define new coordinates
\be
 \left(\begin{array}{c} u_{21}\\ v_{21} \end{array}\right) = \left(\begin{array}{c} u_{01}/(v_{01}-qt) \\ v_{01} -qt \end{array}\right)\quad \Rightarrow \quad
 \left(\begin{array}{c} u_{01}\\ v_{01} \end{array}\right) = \left(\begin{array}{c} u_{21}v_{21} \\ v_{21}+qt \end{array}\right).
\ee
The equations in these coordinates become
\bse\label{eqn21}
\bea
\left\{\begin{array}{rcl} \wb{u} & = & \dfrac{u_{21} v_{21} \left(v_{21} + qt\right) \left(q u_{21} v_{21} - qt - u_{21}^2 v_{21}^3t - q u_{21}^2 v_{21}^2t^2\right)}{q\left(u_{21}v_{21}-t\right){}^2}\\ 
					   \wb{v} & = & \dfrac{u_{21} v_{21}-t}{u_{21}^2 v_{21}^2 \left(v_{21} + qt\right)} \end{array} \right. 
\eea
and
\bea					   
					     \left\{\begin{array}{rcl} \ub{u} & = & \dfrac{1}{u_{21} \left(v_{21} + qt\right){}^2}\\ 
         					   \ub{v} & = & u_{21} \left(v_{21} + qt\right) \left(1 - q^2 u_{21} v_{21}^2t - 2q^3 u_{21} v_{21}t^2 - q^4 u_{21}t^3 \right) \end{array}\right.
\eea
\ese
where the base point $b_{2}$ is replaced by the exceptional line $E_{2}$ defined by $v_{21}=0$. There are no new base points in this chart.

To look in the other chart, define new coordinates
\be
 \left(\begin{array}{c} u_{22}\\ v_{22} \end{array}\right) = \left(\begin{array}{c} u_{01} \\ (v_{01}-qt)/u_{01}\end{array}\right)\quad \Rightarrow \quad
 \left(\begin{array}{c} u_{01}\\ v_{01} \end{array}\right) = \left(\begin{array}{c} u_{22} \\ u_{22}v_{22}+qt \end{array}\right).
\ee

The equations in these coordinates become
\bse\label{eqn22}
\bea
\left\{\begin{array}{rcl} \wb{u} & = & \dfrac{u_{22} \left(u_{22} v_{22} + qt \right) \left( q u_{22} - qt - u_{22}^3 v_{22}t - q u_{22}^2t^2 \right)}{q \left( u_{22}-t \right){}^2}\\ 
					   \wb{v} & = & \dfrac{u_{22}-t}{u_{22}^2 \left(u_{22} v_{22} + qt\right)} \end{array} \right. 
					   \eea
					   and
					   \bea         \left\{\begin{array}{rcl} \ub{u} & = &\dfrac{v_{22}}{\left(u_{22} v_{22} + qt \right){}^2}\\ &&\\
         					   \ub{v} & = &\dfrac{\left(u_{22} v_{22} + qt \right) \left(v_{22}  - q^2 u_{22}^2 v_{22}^2t - 2 q^3 u_{22} v_{22}t^2 - q^4t^3 \right)}{v_{22}^2} \end{array}\right.
\eea
\ese
where $E_2$ is defined by $u_{22}=0$. 
There are no new base points in this chart.


\subsection{Blow up of base point $b_{3}$}\label{a:b3}
Define new coordinates
\be
 \left(\begin{array}{c} u_{31}\\ v_{31} \end{array}\right) = \left(\begin{array}{c} u_{02}/v_{02} \\ v_{02} \end{array}\right)\quad \Rightarrow \quad
 \left(\begin{array}{c} u_{02}\\ v_{02} \end{array}\right) = \left(\begin{array}{c} u_{31}v_{31} \\ v_{31} \end{array}\right).
\ee
In terms of the original variables, this is
\be
 \left(\begin{array}{c} u_{31}\\ v_{31} \end{array}\right) = \left(\begin{array}{c} 1/(uv) \\ v \end{array}\right) \quad \Rightarrow \quad
 \left(\begin{array}{c} u \\ v \end{array}\right) = \left(\begin{array}{c} 1/(u_{31}v_{31}) \\ v_{31} \end{array}\right).
\ee
The equations in these coordinates become
\bse\label{eqn31}
\bea
\left\{\begin{array}{rcl} \wb{u} & = & \dfrac{v_{31} \left(q u_{31} - q u_{31}^2 v_{31}t - t \right)}{q u_{31} \left(1- u_{31} v_{31}t\right){}^2}\\ &&\\
					   \wb{v} & = & u_{31} \left(1- u_{31} v_{31}t\right) \end{array} \right. 
\eea
and
\bea					   
					        \left\{\begin{array}{rcl} \ub{u} & = & \dfrac{u_{31} \left( v_{31}-qt \right)}{v_{31}}\\ 
         					   \ub{v} & = & \dfrac{u_{31} v_{31}-q^2 v_{31}t - q u_{31}t}{u_{31}^2 \left(v_{31}-qt\right){}^2} \end{array}\right.
\eea
\ese
where the base point $b_{3}$ is replaced by the exceptional line $E_{3}$ defined by $v_{31}=0$.
There is a new base point at
\be
 b_{5}: \left(\begin{array}{c} u_{31} \\ v_{31} \end{array}\right) = \left(\begin{array}{c} 0 \\ 0 \end{array}\right).
\ee

To look in the other chart, define new coordinates
\be
 \left(\begin{array}{c} u_{32}\\ v_{32} \end{array}\right) = \left(\begin{array}{c} u_{02} \\ v_{02}/u_{02} \end{array}\right)\quad \Rightarrow \quad
 \left(\begin{array}{c} u_{02}\\ v_{02} \end{array}\right) = \left(\begin{array}{c} u_{32} \\ u_{32}v_{32} \end{array}\right).
\ee
In terms of the original variables, this is
\be
 \left(\begin{array}{c} u_{32}\\ v_{32} \end{array}\right) = \left(\begin{array}{c} 1/u \\ uv \end{array}\right)\quad \Rightarrow \quad
 \left(\begin{array}{c} u \\ v \end{array}\right) = \left(\begin{array}{c} 1/u_{32} \\ u_{32}v_{32} \end{array}\right).
\ee
The equations now become
\bse\label{eqn32}
\bea
&&	\left\{\begin{array}{rcl} \wb{u} & = &\dfrac{u_{32} v_{32} \left(q - q u_{32}t - v_{32}t\right)}{q \left(1 - u_{32}t\right){}^2},\\ 
					   \wb{v} & = &\dfrac{1- u_{32}t}{v_{32}}, \end{array} \right. 
\eea
and
\bea					   				    
         \left\{\begin{array}{rcl} \ub{u} & = & \dfrac{u_{32} v_{32} - qt}{u_{32} v_{32}^2},\\ 
         					   \ub{v} & = & \dfrac{v_{32} \left( u_{32} v_{32}-q^2 u_{32} v_{32}^2t - qt \right)}{\left(u_{32} v_{32} - qt\right){}^2}. \end{array}\right.
\eea
\ese
where $E_3$ is now given by $u_{32}=0$. There are no further base points in this chart.


\subsection{Blow up of base point $b_{4}$}\label{a:b4}
Define new coordinates
\be
 \left(\begin{array}{c} u_{41}\\ v_{41} \end{array}\right) = \left(\begin{array}{c} u_{03}/v_{03} \\ v_{03} \end{array}\right)\quad \Rightarrow \quad
 \left(\begin{array}{c} u_{03}\\ v_{03} \end{array}\right) = \left(\begin{array}{c} u_{41}v_{41} \\ v_{41} \end{array}\right).
\ee
In terms of the original variables, this is
\be
 \left(\begin{array}{c} u_{41}\\ v_{41} \end{array}\right) = \left(\begin{array}{c} uv \\ 1/v \end{array}\right)\quad \Rightarrow \quad
 \left(\begin{array}{c} u\\ v \end{array}\right) = \left(\begin{array}{c} u_{41}v_{41} \\ 1/v_{41} \end{array}\right).
\ee
The equations in these coordinates become
\bse\label{eqn41}
\bea
&&	\left\{\begin{array}{rcl} \wb{u} & = & \dfrac{u_{41} \left(q u_{41} v_{41} - qt - u_{41}^2 v_{41}t\right)}{q \left(u_{41} v_{41} - t\right){}^2}\\ 
					   \wb{v} & = & \dfrac{u_{41} v_{41} - t}{u_{41}^2 v_{41}} \end{array} \right.
		\eea
		and
		\bea 
         \left\{\begin{array}{rcl} \ub{u} & = &\dfrac{1- q v_{41}t}{u_{41}}\\ 
         					  \ub{v} & = & \dfrac{u_{41} v_{41} \left(1 - q^2 u_{41}t - q v_{41}t\right)}{\left(1 - q v_{41}t\right){}^2} \end{array}\right.
\eea
\ese
where $b_4$ is replaced by the exceptional line $E_{4}$, which is defined by $v_{41}=0$.
There are no new base points in this chart.

To look in the other chart, define new coordinates
\be
 \left(\begin{array}{c} u_{42}\\ v_{42} \end{array}\right) = \left(\begin{array}{c} u_{03} \\ v_{03}/u_{03} \end{array}\right)\quad \Rightarrow \quad
 \left(\begin{array}{c} u_{03}\\ v_{03} \end{array}\right) = \left(\begin{array}{c} u_{42} \\ u_{42}v_{42} \end{array}\right).
\ee
In terms of the original variables, this is
\be
 \left(\begin{array}{c} u_{42}\\ v_{42} \end{array}\right) = \left(\begin{array}{c} u \\ 1/(uv) \end{array}\right)\quad \Rightarrow \quad
 \left(\begin{array}{c} u\\ v \end{array}\right) = \left(\begin{array}{c} u_{42} \\ 1/(u_{42}v_{42}) \end{array}\right).
\ee
The equations now become
\bse\label{eqn42}
\bea
&& \left\{\begin{array}{rcl} \wb{u} & = &\dfrac{q u_{42} v_{42} - q v_{42}t - u_{42}t }{q v_{42}^2 \left(u_{42} - t\right){}^2}\\ 
					\wb{v} & = &\dfrac{v_{42} \left(u_{42} - t\right)}{u_{42}}\end{array} \right.  
\eea
and
\bea         \left\{\begin{array}{rcl} \ub{u} & = &v_{42} \left(1 - q u_{42} v_{42}t\right)\\
         					   \ub{v} & = & \dfrac{u_{42} \left(v_{42}-q^2t - q u_{42} v_{42}^2t\right)}{v_{42} \left(1 - q u_{42} v_{42}t\right){}^2} \end{array}\right.
\eea
\ese
where $E_{4}$ is defined by $u_{42}=0$. 
There is a new base point in this chart at
\be
 b_{6}: \left(\begin{array}{c} u_{42} \\ v_{42} \end{array}\right) = \left(\begin{array}{c} 0 \\ 0 \end{array}\right).
\ee


\subsection{Blow up of base point $b_{5}$}\label{a:b5}
Define new coordinates
\be
 \left(\begin{array}{c} u_{51}\\ v_{51} \end{array}\right) = \left(\begin{array}{c} u_{31}/v_{31} \\ v_{31} \end{array}\right)\quad \Rightarrow \quad
 \left(\begin{array}{c} u_{31}\\ v_{31} \end{array}\right) = \left(\begin{array}{c} u_{51}v_{51} \\ v_{51} \end{array}\right).
\ee
In terms of the original variables, this is
\be
 \left(\begin{array}{c} u_{51}\\ v_{51} \end{array}\right) = \left(\begin{array}{c} 1/(uv^2) \\ v \end{array}\right)\quad \Rightarrow \quad
 \left(\begin{array}{c} u \\ v \end{array}\right) = \left(\begin{array}{c} 1/(u_{51}v_{51}^2) \\ v_{51} \end{array}\right).
\ee
The equations in these coordinates become
\bse\label{eqn51}
\bea
&&	\left\{\begin{array}{rcl} \wb{u} & = & \dfrac{q u_{51} v_{51} - q u_{51}^2 v_{51}^3t - t}{q u_{51} \left(1 - u_{51} v_{51}^2t\right){}^2}\\ 
					   \wb{v} & = & u_{51} v_{51} \left(1 - u_{51} v_{51}^2t\right) \end{array} \right. 
					   \eea
					   and
					   \bea
         \left\{\begin{array}{rcl} \ub{u} & = & u_{51} \left(v_{51} - qt\right)\\ 
         					   \ub{v} & = & \dfrac{u_{51} v_{51} - q^2t - q u_{51}t}{u_{51}^2 v_{51} \left(v_{51} - qt\right){}^2} \end{array}\right.
\eea
\ese
where $b_{5}$ is replaced by the exceptional line $E_{5}$ defined by $v_{51}=0$. There is a new base point  at
\be
  b_{7}: \left(\begin{array}{c} u_{51} \\ v_{51} \end{array}\right) = \left(\begin{array}{r} -q \\ 0 \end{array}\right).
\ee

To look in the other chart, define new coordinates
\be
 \left(\begin{array}{c} u_{52}\\ v_{52} \end{array}\right) = \left(\begin{array}{c} u_{31} \\ v_{31}/u_{31} \end{array}\right)\quad \Rightarrow \quad
 \left(\begin{array}{c} u_{31}\\ v_{31} \end{array}\right) = \left(\begin{array}{c} u_{52}\\ u_{52}v_{52} \end{array}\right).
\ee
In terms of the original variables, this is
\be
 \left(\begin{array}{c} u_{52}\\ v_{52} \end{array}\right) = \left(\begin{array}{c} 1/(uv) \\ uv^2 \end{array}\right)\quad \Rightarrow \quad
 \left(\begin{array}{c} u \\ v \end{array}\right) = \left(\begin{array}{c} 1/(u_{52}^2v_{52}) \\ u_{52}v_{52} \end{array}\right).
\ee

The equations in these coordinates become
\bse\label{eqn52}
\bea
&&	\left\{\begin{array}{rcl} \wb{u} & = & \dfrac{ v_{52} \left(q u_{52} - q u_{52}^3 v_{52}t - t\right)}{q \left(1 - u_{52}^2 v_{52}t\right){}^2}\\ 
					   \wb{v} & = & u_{52} \left(1 - u_{52}^2 v_{52}t\right) \end{array} \right. 
					   \eea
					   and
					   \bea 
         \left\{\begin{array}{rcl} \ub{u} & = & \dfrac{u_{52} v_{52} - qt}{v_{52}}\\ 
         					   \ub{v} & = & \dfrac{u_{52} v_{52} - q^2 v_{52}t - qt}{u_{52} \left(u_{52} v_{52} - qt\right){}^2} \end{array}\right.
\eea
\ese
whereas $E_5$ is now $u_{52}=0$. There are no further base points appearing here.


\subsection{Blow up of base point $b_{6}$}\label{a:b6}
Define new coordinates
\be
 \left(\begin{array}{c} u_{61}\\ v_{61} \end{array}\right) = \left(\begin{array}{c} u_{42}/v_{42} \\ v_{42} \end{array}\right)\quad \Rightarrow \quad
 \left(\begin{array}{c} u_{42}\\ v_{42} \end{array}\right) = \left(\begin{array}{c} u_{61}v_{61} \\ v_{61} \end{array}\right).
\ee
In terms of the original variables, this is
\be
 \left(\begin{array}{c} u_{61}\\ v_{61} \end{array}\right) = \left(\begin{array}{c} u^2v \\ 1/(uv) \end{array}\right)\quad \Rightarrow \quad
 \left(\begin{array}{c} u \\ v \end{array}\right) = \left(\begin{array}{c} u_{61}v_{61} \\ 1/(u_{61}v_{61}^2) \end{array}\right).
\ee
The equations in these new coordinates become
\bse\label{eqn61}
\bea
&&\left\{\begin{array}{rcl} \wb{u} & = & \dfrac{q u_{61} v_{61} - qt - u_{61}t}{q v_{61} \left(u_{61} v_{61} - t\right){}^2}\\ 
				   \wb{v} & = & \dfrac{u_{61} v_{61} - t}{u_{61}} \end{array} \right. 
				   \eea
				   and
				   \bea
         \left\{\begin{array}{rcl} \ub{u} & = & v_{61} \left(1 - q u_{61} v_{61}^2t\right)\\ 
         					   \ub{v} & = & \dfrac{u_{61} \left(v_{61} - q^2t - q u_{61} v_{61}^3t\right)}{\left(1 - q u_{61} v_{61}^2t\right){}^2} \end{array}\right.
\eea
\ese
where the base point $b_{6}$ is replaced by the exceptional line $E_{6}$ defined by $v_{61}=0$.  There is a new base point  at
\be
  b_{8}: \left(\begin{array}{c} u_{61} \\ v_{61} \end{array}\right) = \left(\begin{array}{r} -q \\ 0 \end{array}\right).
\ee

To look in the other chart, define new coordinates
\be
 \left(\begin{array}{c} u_{62}\\ v_{62} \end{array}\right) = \left(\begin{array}{c} u_{42} \\ v_{42}/u_{42} \end{array}\right)\quad \Rightarrow \quad
 \left(\begin{array}{c} u_{42}\\ v_{42} \end{array}\right) = \left(\begin{array}{c} u_{62} \\ u_{62}v_{62} \end{array}\right).
\ee
In terms of the original variables, this is
\be
 \left(\begin{array}{c} u_{62}\\ v_{62} \end{array}\right) = \left(\begin{array}{c} u \\ 1/(u^2v) \end{array}\right)\quad \Rightarrow \quad
 \left(\begin{array}{c} u \\ v \end{array}\right) = \left(\begin{array}{c} u_{62} \\ 1/(u_{62}^2v_{62}) \end{array}\right).
\ee
The equations in these new coordinates become
\bse\label{eqn62}
\bea
&&	\left\{\begin{array}{rcl} \wb{u} & = & \dfrac{q u_{62} v_{62} - q v_{62}t - t}{q u_{62} v_{62}^2 \left(u_{62} - t\right){}^2}\\ 
					   \wb{v} & = & v_{62} \left(u_{62} - t\right) \end{array} \right. 
					   \eea
					   and
					   \bea
					            \left\{\begin{array}{rcl} \ub{u} & = & u_{62} v_{62} \left(1 - q u_{62}^2 v_{62}t\right)\\ 
         					   \ub{v} & = & \dfrac{\left(u_{62} v_{62} - q^2t - q u_{62}^3 v_{62}^2t\right)}{v_{62} \left(1 - q u_{62}^2 v_{62}t\right){}^2} \end{array}\right.
\eea
\ese
where $E_{6}$ is now defined by $u_{62}=0$. There are no further base points in this chart.


\subsection{Blow up of base point $b_{7}$}\label{a:b7}
Define new coordinates
\be
 \left(\begin{array}{c} u_{71}\\ v_{71} \end{array}\right) = \left(\begin{array}{c} (u_{51}+q)/v_{51} \\ v_{51} \end{array}\right)\quad \Rightarrow \quad
 \left(\begin{array}{c} u_{51}\\ v_{51} \end{array}\right) = \left(\begin{array}{c} u_{71}v_{71}-q \\ v_{71} \end{array}\right).
\ee
In terms of the original variables, this is
\be
 \left(\begin{array}{c} u_{71}\\ v_{71} \end{array}\right) = \left(\begin{array}{c} (1+quv^2)/(uv^3) \\ v \end{array}\right) \quad \Rightarrow \quad
 \left(\begin{array}{c} u \\ v \end{array}\right) = \left(\begin{array}{c} 1/[(u_{71}v_{71}-q)v_{71}^2] \\ v_{71} \end{array}\right).
\ee
The equations now become
\bea
	\left\{\begin{array}{rcl} \wb{u} & = & \dfrac{q^2 v_{71} - q u_{71} v_{71}^2 + q^3 v_{71}^3t - 2 q^2 u_{71} v_{71}^4t + q u_{71}^2 v_{71}^5t + t}
									{q \left(q-u_{71}v_{71}\right) \left(1 + q v_{71}^2t - u_{71} v_{71}^3t\right){}^2}\\ 
					   \wb{v} & = & v_{71} \left(u_{71} v_{71}-q\right) \left(1 + q v_{71}^2t - u_{71} v_{71}^3t\right), \end{array} \right.
\eea
and
\bea
         \left\{\begin{array}{rcl} \ub{u} & = & -\left(q - u_{71} v_{71}\right) \left(v_{71} - qt\right)\\ 
         					   \ub{v} & = & \dfrac{-q + u_{71} v_{71} - q u_{71}t}{\left(u_{71} v_{71}-q\right){}^2 \left(v_{71} - qt\right){}^2} \end{array}\right.\label{eqn71}
\eea
where the base point $b_{7}$ is now replaced by the exceptional line $E_{7}$ defined by $v_{71}=0$. There are no further base points in this chart.

To look in the other chart, define new coordinates
\be
 \left(\begin{array}{c} u_{72}\\ v_{72} \end{array}\right) = \left(\begin{array}{c} u_{51}+q \\ v_{51}/(u_{51}+q) \end{array}\right)\quad \Rightarrow \quad
 \left(\begin{array}{c} u_{51}\\ v_{51} \end{array}\right) = \left(\begin{array}{c} u_{72}-q \\ u_{72}v_{72} \end{array}\right).
\ee
In terms of the original variables, this is
\be
 \left(\begin{array}{c} u_{72}\\ v_{72} \end{array}\right) = \left(\begin{array}{c} (1+quv^2)/(uv^2) \\ uv^3/(1+quv^2) \end{array}\right)\quad \Rightarrow \quad
 \left(\begin{array}{c} u \\ v \end{array}\right) = \left(\begin{array}{c} 1/[(u_{72}-q)u_{72}^2v_{72}^2] \\ u_{72}v_{72} \end{array}\right).
\ee
The equations in these coordinates become
\bea
\left\{\begin{array}{rcl} \wb{u} & = &\dfrac{q^2 u_{72} v_{72} - q u_{72}^2 v_{72} + q^3 u_{72}^3 v_{72}^3t - 2 q^2 u_{72}^4 v_{72}^3t + q u_{72}^5 v_{72}^3t + t}
									{q\left(q-u_{72}\right) \left(1 + q u_{72}^2 v_{72}^2t - u_{72}^3 v_{72}^2t\right){}^2}\\ 
					   \wb{v} & = & u_{72} v_{72} \left(u_{72}-q\right) \left(1 + q u_{72}^2 v_{72}^2t - u_{72}^3 v_{72}^2t\right) \end{array} \right.
\eea
and
\bea
         \left\{\begin{array}{rcl} \ub{u} & = & -\left(q-u_{72}\right) \left(u_{72} v_{72} - qt\right)\\ 
         					   \ub{v} & = &\dfrac{-q v_{72} + u_{72} v_{72} - qt}{v_{72} \left(u_{72} - q\right){}^2 \left(u_{72} v_{72} - qt\right){}^2} \end{array}\right.\label{eqn72}
\eea
where now $E_7$ is defined by $u_{72}=0$. There are no further base points appearing here.


\subsection{Blow up of base point $b_{8}$}\label{a:b8}
Define new coordinates
\be
 \left(\begin{array}{c} u_{81}\\ v_{81} \end{array}\right) = \left(\begin{array}{c} (u_{61}+q)/v_{61} \\ v_{61} \end{array}\right)\quad \Rightarrow \quad
 \left(\begin{array}{c} u_{61}\\ v_{61} \end{array}\right) = \left(\begin{array}{c} u_{81}v_{81}-q \\ v_{81} \end{array}\right).
\ee
In terms of the original variables, this is
\be
 \left(\begin{array}{c} u_{81}\\ v_{81} \end{array}\right) = \left(\begin{array}{c} (u^2v+q)uv \\ 1/(uv) \end{array}\right)\quad \Rightarrow \quad
 \left(\begin{array}{c} u \\ v \end{array}\right) = \left(\begin{array}{c} (u_{81}v_{81}-q)v_{81} \\ 1/[(u_{81}v_{81}-q)v_{81}^2] \end{array}\right).
\ee
The equations become
\bea
\left\{\begin{array}{rcl} \wb{u} & = & \dfrac{q u_{81} v_{81} - q^2 - u_{81}t}{q \left(q v_{81} - u_{81} v_{81}^2 + t\right){}^2}\\ 
				        \wb{v} & = & \dfrac{q v_{81} - u_{81} v_{81}^2 + t}{q - u_{81} v_{81}t} \end{array} \right. 
				        \eea
				        and
				        \bea
         \left\{\begin{array}{rcl} \ub{u} & = & v_{81} \left(1 + q^2 v_{81}^2t - q u_{81} v_{81}^3t\right)\\ 
         					   \ub{v} & = & \dfrac{\left(u_{81} v_{81} - q\right) \left(v_{81} + q^2 v_{81}^3t - q^2t - q u_{81} v_{81}^4t\right)}{\left(1 + q^2 v_{81}^2t - q u_{81}v_{81}^3t\right){}^2} \end{array}\right.\label{eqn81}
\eea
where the base point $b_{8}$ is now replaced by the exceptional line $E_{8}$ defined by $v_{81}=0$. There are no further base points in this chart.

To look in the other chart, define new coordinates
\be
 \left(\begin{array}{c} u_{82}\\ v_{82} \end{array}\right) = \left(\begin{array}{c} u_{61}+q \\ v_{61}/(u_{61}+q) \end{array}\right)\quad \Rightarrow  \quad
 \left(\begin{array}{c} u_{61}\\ v_{61} \end{array}\right) = \left(\begin{array}{c} u_{82}-q \\ u_{82}v_{82} \end{array}\right).
\ee
In terms of the original variables, this is
\be
 \left(\begin{array}{c} u_{82}\\ v_{82} \end{array}\right) = \left(\begin{array}{c} u^2v+q \\ 1/[(u^2v+q)uv] \end{array}\right)\quad \Rightarrow \quad
 \left(\begin{array}{c} u \\ v \end{array}\right) = \left(\begin{array}{c} (u_{82}-q)u_{82}v_{82} \\ 1/[(u_{82}-q)u_{82}^2v_{82}^2] \end{array}\right).
\ee
The equations in these coordinates become
\bea
\left\{\begin{array}{rcl} \wb{u} & = & \dfrac{q u_{82} v_{82} - q^2 v_{82} - t}{q v_{82} \left(q u_{82} v_{82} - u_{82}^2 v_{82} + t\right){}^2}\\ 
					\wb{v} & = & \dfrac{q u_{82} v_{82} - u_{82}^2 v_{82} + t}{q - u_{82}} \end{array} \right. 
					\eea
					and
					\bea         \left\{\begin{array}{rcl} \ub{u} & = & u_{82} v_{82} \left(1 + q^2 u_{82}^2 v_{82}^2t - q u_{82}^3 v_{82}^2t\right)\\ &&\\
         					   \ub{v} & = &\dfrac{\left(u_{82} - q\right) \left(u_{82}v_{82} + q^2 u_{82}^3 v_{82}^3t - q^2t - q u_{82}^4 v_{82}^3t\right)}
					   				{\left(1 + q^2 u_{82}^2 v_{82}^2t - qu_{82}^3 v_{82}^2t\right){}^2} \end{array}\right.\label{eqn82}
\eea
where $E_8$ is now defined by $u_{82}=0$. There are no further base points in this chart.

\section{Dynamics of solutions near divisors $D_{i}$}\label{appendix_divisors}
Defining $A_{ij}:=2(D_{i},D_{j})/(D_{j},D_{j})$, we can express the intersection information between $D_i$ and $D_j$ in a generalised Cartan matrix $A:=(A_{ij})_{i,j=1}^{8}$. For $\{D_j\}_{j=1}^8$ defined in \S \ref{section_divisors}, we have 
\benn
A=
 \left(\begin{array}{cccccccc} 	2 & -1 & 0 & 0 & 0 & 0 & 0 & -1 \\
						-1 & 2 & -1 & 0 & 0 & 0 & 0 & 0 \\
						0 & -1 & 2 & -1 & 0 & 0 & 0 & 0 \\
						0 & 0 & -1 & 2 & -1 & 0 & 0 & 0 \\
						0 & 0 & 0 & -1 & 2 & -1 & 0 & 0 \\
						0 & 0 & 0 & 0 & -1 & 2 & -1 & 0 \\
						0 & 0 & 0 & 0 & 0 & -1 & 2 & -1 \\
						-1 & 0 & 0 & 0 & 0 & 0 & -1 & 2\end{array}\right).
\eenn
\subsection{Symmetries of the system}
\begin{figure}[H]
\begin{center}
\begin{tikzpicture}[start chain=circle placed {at=(\tikzchaincount*180:1)}]
  \foreach \i in {1,...,2}
    \node [draw,thick,circle,on chain,join] {\tiny $F_{\i}$};
   \node[above] (0,0) {$\infty$};
\end{tikzpicture}
\caption{Dynkin diagram corresponding to the symmetry group.}
\label{fig_orth}
 \end{center}
 \end{figure}
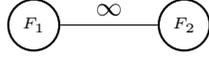

We find the symmetries of the system by constructing vectors orthogonal to the components $D_{1},\dots,D_{8}$, and defining corresponding actions which leave this set invariant.
A vector $\alpha\in$ Pic$(\mathcal S)$ is given by the linear combination
\be
 \al := \al_{u}H_{u}+\al_{v}H_{v}+\sum_{i=1}^{8}\al_{i}E_{i}
\ee
and its intersection with each $D_j$ is given by
\begin{align*}
 (\al,D_{1}) &= \al_{u}+\al_{1}+\al_{3},\\
 (\al,D_{2}) &= -\al_{3}+\al_{5},\\
 (\al,D_{3}) &= -\al_{5}+\al_{7},\\
 (\al,D_{4}) &= \al_{v}+\al_{3}+\al_{5},\\
 (\al,D_{5}) &= \al_{u}+\al_{4}+\al_{6},\\
 (\al,D_{6}) &= -\al_{6}+\al_{8},\\
 (\al,D_{7}) &= -\al_{4}+\al_{6},\\
 (\al,D_{8}) &= \al_{v}+\al_{2}+\al_{4}.
\end{align*}
For orthogonality to be satisfied, it follows that 
\bea
 \al_{3} = \al_{5} = \al_{7} & =: & a,\\
 \al_{4} = \al_{6} = \al_{8} & =: & b,
\eea
where $a$ and $b$ are arbitrary, and hence
\bea
 \al_{u} & = & -2b,\\
 \al_{v} & = & -2a,\\
 \al_{1} & = & 2b-a,\\
 \al_{2} & = & 2a-b.
\eea
Thus $\al$ becomes
\begin{align}
 \al &= a(-2H_{v}-E_{1}+2E_{2}+E_{3}+E_{5}+E_{7}) + b(-2H_{u}+2E_{1}-E_{2}+E_{4}+E_{6}+E_{8})\nn\\
  & =: aF_{1}+bF_{2}
\end{align}
where, if we define $B_{i,j}:=2(F_{i},F_{j})/(F_{j},F_{j})$, we find the generalised Cartan matrix $B:=(B_{ij})_{i,j=1}^{2}$ given by
\be
B=
 \left(\begin{array}{rr} 	
 					2 & -2  \\
					-2 & 2
\end{array}\right).
\ee
This leads to the Dynkin diagram shown in Figure \ref{fig_orth}, which corresponds to the root lattice $A_{1}^{(1)}$.
 
For each vector $F_{i}$, we define an action on an element $x\in \textrm{Pic}(X)$ as follows:
\bea
 w_{F_{i}}(x) & := & x - 2\dfrac{(x,F_{i})}{(F_{i},F_{i})}F_{i}\nn\\
 		    & = & x + \dfrac{1}{4}(x,F_{i})F_{i}.
\eea

The action of $w_{F_{1}}$ on Pic($X$) is
\be
 w_{F_{1}}\left( \begin{array}{c} H_{u}\\ H_{v}\\ E_{1}\\ E_{2}\\ E_{3}\\ E_{4}\\ E_{5}\\ E_{6}\\ E_{7}\\ E_{8} \end{array} \right)
  = \left(
\begin{array}{cccccccccc}
 1 & 1 & 1/2 & -1 & -1/2 & 0 & -1/2 & 0 & -1/2 & 0\\
 0 & 1 & 0 & 0 & 0 & 0 & 0 & 0 & 0 & 0\\
 0 & -1/2 & 3/4 & 1/2 & 1/4 & 0 & 1/4 & 0 & 1/4 & 0\\
 0 & 1 & 1/2 & 0 & -1/2 & 0 & -1/2 & 0 & -1/2 & 0\\
 0 & 1/2 & 1/4 & -1/2 & 3/4 & 0 & -1/4 & 0 & -1/4 & 0\\
 0 & 0 & 0 & 0 & 0 & 1 & 0 & 0 & 0 & 0\\
 0 & 1/2 & 1/4 & -1/2 & -1/4 & 0 & 3/4 & 0 & -1/4 & 0\\
 0 & 0 & 0 & 0 & 0 & 0 & 0 & 1 & 0 & 0\\
 0 & 1/2 & 1/4 & -1/2 & -1/4 & 0 & -1/4 & 0 & 3/4 & 0\\
 0 & 0 & 0 & 0 & 0 & 0 & 0 & 0 & 0 & 1
\end{array}
\right)
      \left( \begin{array}{c} H_{u}\\ H_{v}\\ E_{1}\\ E_{2}\\ E_{3}\\ E_{4}\\ E_{5}\\ E_{6}\\ E_{7}\\ E_{8} \end{array} \right),
\ee
and the action of $w_{F_{2}}$ on Pic($X$) is
\be
 w_{F_{2}}\left( \begin{array}{c} H_{u}\\ H_{v}\\ E_{1}\\ E_{2}\\ E_{3}\\ E_{4}\\ E_{5}\\ E_{6}\\ E_{7}\\ E_{8} \end{array} \right)
  = \left(
\begin{array}{cccccccccc}
 1 & 0 & 0 & 0 & 0 & 0 & 0 & 0 & 0 & 0 \\
 1 & 1 & -1 & 1/2 & 0 & -1/2 & 0 & -1/2 & 0 & -1/2 \\
 1 & 0 & 0 & 1/2 & 0 & -1/2 & 0 & -1/2 & 0 & -1/2 \\
 -1/2 & 0 & 1/2 & 3/4 & 0 & 1/4 & 0 & 1/4 & 0 & 1/4 \\
 0 & 0 & 0 & 0 & 1 & 0 & 0 & 0 & 0 & 0 \\
 1/2 & 0 & -1/2 & 1/4 & 0 & 3/4 & 0 & -1/4 & 0 & -1/4 \\
 0 & 0 & 0 & 0 & 0 & 0 & 1 & 0 & 0 & 0 \\
 1/2 & 0 & -1/2 & 1/4 & 0 & -1/4 & 0 & 3/4 & 0 & -1/4 \\
 0 & 0 & 0 & 0 & 0 & 0 & 0 & 0 & 1 & 0 \\
1/2 & 0 & -1/2 & 1/4 & 0 & -1/4 & 0 & -1/4 & 0 & 3/4
\end{array}
\right)
      \left( \begin{array}{c} H_{u}\\ H_{v}\\ E_{1}\\ E_{2}\\ E_{3}\\ E_{4}\\ E_{5}\\ E_{6}\\ E_{7}\\ E_{8} \end{array} \right).
\ee

Note that the action of $w_{F_{i}}$ on each of the $D_{i}$ is the identity, i.e., 
\be
 w_{F_{i}}(D_{j}) = D_{j} \quad \forall i=1,2; \quad j=1,..,8,
\ee
and taking either of these actions to the second power gives the identity, so the $w_{F_{i}}$ are indeed reflections; the span of these reflections forms the affine Weyl group $W$. 
Including a Dynkin automorphism $\sigma: F_{1}\leftrightarrow F_{2}$, we obtain the extended affine Weyl group $\widetilde{W}$. Note that the action $(w_{F_{2}}w_{F_{1}})^4$ has the same effect on the elements of the Picard group as does the 4th power of the mapping \eqref{qP1}. That is, if we denote the operation of forward shift in the discrete time variable by $\varphi$, we have
\be
 \varphi^4(K) = (w_{F_{2}}w_{F_{1}})^4 K,
\ee
for $K$ an element of the Picard group.

In the following, we examine behaviour of solutions near each irreducible component $D_{1},\dots,D_{8}$ of the anti-canonical divisor. We focus on the forward iteration here, as the case of backward iteration is entirely analogous.



\subsection{Behaviour near $D_{1}:=H_{v=0}-E_{1}-E_{3}$}
The component $D_{1}$ is essentially the coordinate axis $v_{01}=v_{02}=0$, where $u_{01}\neq t,u_{02}\neq0$. 
Suppose that at a time $t_{1}$ we are near $D_{1}$, i.e., $v_{01}(t_{1})=v_{02}(t_{1})$ is close to zero.
Expanding the equations \eqref{qp1} for forward iteration in the chart $(u_{41}, v_{41})$, we find to leading order for $v_{01}(t_1)=\epsilon\ll 1$
\be
	\left\{\begin{array}{rcl} \wb{u}_{41}(t_{1}) & = & \dfrac{1}{u_{01}(t_{1})}+{\mathcal O}(\epsilon)  \\ 
					   \wb{v}_{41}(t_{1}) & \sim & \dfrac{u_{01}(t_{1})^2v_{01}(t_{1})}{u_{01}(t_{1}) - t_{1}}={\mathcal O}(\epsilon) \end{array} \right. 
\ee
while for $v_{02}(t_1)=\epsilon$ we have
\be
	\left\{\begin{array}{rcl} \wb{u}_{41}(t_{1}) & = & u_{02}(t_{1})+{\mathcal O}(\epsilon)  \\ 
					   \wb{v}_{41}(t_{1}) & \sim & \dfrac{v_{02}(t_{1})}{u_{02}(t_{1})^2(1 - u_{02}(t_{1})t_{1})}={\mathcal O}(\epsilon) . \end{array} \right. 
\ee
The image lies near the line $v_{41}=0$, that is, near the component $D_{7}$.


\subsection{Behaviour near $D_{2}$}
The component $D_{2}$ is defined in local coordinates by $v_{31}=0$, $u_{32}=0$, where $u_{31}\neq0$.
Suppose that at a time $t_{2}$ we are near this divisor; i.e., $v_{31}(t_{2}),u_{32}(t_{2})$ are close to zero.
Expanding the equations \eqref{qp1} for forward iteration in the chart $(u_{31}, v_{31})$, we find to leading order for $v_{31}(t_2)=\epsilon\ll 1$
\be
\left\{\begin{array}{rcl}
		 \wb{u}_{01}(t_{2}) & \sim & \dfrac{v_{31}(t_{2})(qu_{31}(t_{2}) - t_{2})}{qu_{31}(t_{2})}={\mathcal O}(\ep) \\
 		 \wb{v}_{01}(t_{2}) & = & u_{31}(t_{2})+{\mathcal O}(\ep) \end{array}\right.
\ee
and for $u_{32}(t_2)=\ep$
\be
\left\{\begin{array}{rcl}
		\wb{u}_{01}(t_{2}) & \sim & \dfrac{v_{32}(t_{2})u_{32}(t_{2})(q - v_{32}(t_{2})t_{2})}{q}={\mathcal O}(\ep)\\
 		\wb{v}_{01}(t_{2}) & = & \dfrac{1}{v_{32}(t_{2})}+{\mathcal O}(\ep). \end{array}\right.
\ee
We see that the image near the line $u_{01}=0$,  near the component $D_{8}$.


\subsection{Behaviour near $D_{3}$}
The component $D_{3}$ is the line $E_{5}-E_{7}$ is defined in local coordinates by $v_{51}=0,u_{52}=0$, where $u_{51}\neq-q$.
Suppose that at a time $t_{3}$ we are near this component; i.e., $v_{51}(t_{3})=\ep$, or $u_{52}(t_{3})=\ep$ where $\ep\ll 1$. 
Expanding the equations \eqref{qp1}, we see that in the coordinates of the first chart
\be
\left\{\begin{array}{rcl}
		 \wb{u}_{01}(t_{3}) & = & -\dfrac{1}{qt_{3}u_{51}(t_{3})}+{\mathcal O}(\ep),\\
 		 \wb{v}_{01}(t_{3}) & \sim & u_{51}(t_{3})v_{51}(t_{3})={\mathcal O}(\ep), \end{array}\right.
\ee
while, in the coordinates of the second chart  we obtain
\be
\left\{\begin{array}{rcl}
		 \wb{u}_{01}(t_{3}) & = & -\dfrac{v_{52}(t_{3})}{qt_{3}}+{\mathcal O}(\ep),\\
 		 \wb{v}_{01}(t_{3}) & \sim &  u_{52}(t_{3}) ={\mathcal O}(\ep). \end{array}\right.
\ee
The image lies near the line $v_{01}=0$, or rather the component $D_{1}$.

\subsection{Behaviour near $D_{4}$} 
The component $D_{4}$ is given by $u_{02}=u_{04}=0$, where $v_{02}\neq0$. 
Suppose that at a time $t_{4}$ we are near this component; i.e., $u_{02}(t_{4})=u_{04}(t_{4})=\ep\ll1$.
Expanding the equations for $u_{31}$, $v_{31}$, we find in the first chart
\be
\left\{\begin{array}{rcl}
		 \wb{u}_{31}(t_{4}) & = & -\dfrac{qt_{4}}{v_{02}(t_{4})}+{\mathcal O}(\ep)\\
 		 \wb{v}_{31}(t_{4}) & \sim & \dfrac{u_{02}(t_{4})}{v_{02}(t_{4})}={\mathcal O}(\ep) \end{array}\right.
\ee
while in the coordinates of the second chart  we obtain
\be
\left\{\begin{array}{rcl}
		 \wb{u}_{31}(t_{4}) & = & -qt_{4}v_{04}(t_{4})+{\mathcal O}(\ep)\\
 		 \wb{v}_{31}(t_{4}) & \sim & v_{04}(t_{4})u_{04}(t_{4})={\mathcal O}(\ep). \end{array}\right.
\ee
The image lies near the line $v_{31}=0$, or rather the component $D_{2}$.


\subsection{Behaviour near $D_{5}$} 
The divisor $D_{5}$ is essentially the coordinate axis $v_{03}=v_{04}=0$, where $u_{03}\neq0$.
Suppose that at a time $t_{5}$ we are near this divisor; i.e., $v_{03}(t_{5})=v_{04}(t_{5})=\ep$ is close to zero.
Expanding the equations for $u_{51}$, $v_{51}$, we find for $|v_{03}|\ll 1$
\be
\left\{\begin{array}{rcl}
		 \wb{u}_{51}(t_{5}) & = & -qt_{5}u_{03}(t_{5})+{\mathcal O}(\ep)\\
 		 \wb{v}_{51}(t_{5}) & \sim & \dfrac{v_{03}(t_{5})(u_{03}(t_{5}) - t_{5})}{u_{03}(t_{5})^2}={\mathcal O}(\ep) \end{array}\right.
\ee
while for $|v_{04}\ll1$, we find
\be
\left\{\begin{array}{rcl}
		 \wb{u}_{51}(t_{5}) & = & -\dfrac{qt_{5}}{u_{04}(t_{5})}+{\mathcal O}(\ep)\\
 		 \wb{v}_{51}(t_{5}) & \sim &  \ u_{04}(t_{5})v_{04}(t_{5})(1 - u_{04}(t_{5})t_{5})+{\mathcal O}(\ep) \end{array}\right.
\ee
The image lies near the line $v_{51}=0$, or rather the component $D_{3}$


\subsection{Behaviour near $D_{6}$}
The component $D_{6}$ is defined in local coordinates by $v_{61}=0,u_{62}=0$, where $u_{61}\neq-q$.
Suppose that at a time $t_{6}$ we are near this component; i.e., $v_{61}(t_{6})$, $u_{62}(t_{6})$ are close to zero. 
Expanding the equations for forward iteration in chart 2, we see that for $|v_{61}|=\ep\ll1$
\be
\left\{\begin{array}{rcl}
		 \wb{u}_{02}(t_{6}) & \sim & -\dfrac{qv_{61}(t_{6})}{t_{6}(q+u_{61}(t_{6}))}={\mathcal O}(\ep) ,\\
 		 \wb{v}_{02}(t_{6}) & = & -\dfrac{1}{t_{6}u_{61}(t_{6})}+{\mathcal O}(\ep), \end{array}\right.
\ee
and for $|u_{62}|=\ep\ll 1$,
\be
\left\{\begin{array}{rcl}
		 \wb{u}_{02}(t_{6}) & \sim & -\dfrac{qv_{62}(t_{6})^2u_{62}(t_{6})}{t_{6}(qv_{62}(t_{6})+1)}={\mathcal O}(\ep)  ,\\
 		 \wb{v}_{02}(t_{6}) & = & -\dfrac{v_{62}(t_{6})}{t_{6}}+{\mathcal O}(\ep). \end{array}\right.
\ee
We see that this is near the line $u_{02}=0$, or rather the component $D_{4}$.

\subsection{Behaviour near $D_{7}$}
The component $D_{7}$ is the line $E_{4}-E_{6}$ that arises under the first blow-up at the point $b_{4}$ in the region where $u$ is finite, $v$ is infinite. It is defined in local coordinates by $v_{41}=0$, $u_{42}=0$, where $v_{42}\neq0$.
Suppose that at a time $t_{7}$ we are near this component; i.e., $v_{41}(t_{7})=\ep$, $u_{42}(t_{7})=\ep$, where $\ep\ll 1$. 
Expanding the equations in chart 3, we see that in the coordinates of the first chart
\be
\left\{\begin{array}{rcl}
		 \wb{u}_{03}(t_{7}) & \sim & -t_{7}u_{41}(t_{7})+{\mathcal O}(\ep) \\
 		 \wb{v}_{03}(t_{7}) & = & - t_{7}u_{41}(t_{7})^2v_{41}(t_{7})={\mathcal O}(\ep)\end{array}\right.
\ee
and in the coordinates of the second chart
\be
\left\{\begin{array}{rcl}
		 \wb{u}_{03}(t_{7}) & = & -\dfrac{t_{7}}{v_{42}(t_{7})}+{\mathcal O}(\ep) ,\\
 		 \wb{v}_{03}(t_{7}) & \sim & -\dfrac{t_{7}u_{42}(t_{7})}{v_{42}(t_{7})}={\mathcal O}(\ep). \end{array}\right.
\ee
The image lies near the line $v_{03}=0$, or rather the component $D_{5}$.


\subsection{Behaviour near $D_{8}$} 
The component $D_{8}$ is essentially the coordinate axis $u_{01}=u_{03}=0$ where $v_{01}\neq qt,v_{03}\neq0$.
Suppose that at a time $t_{8}$ we are near this component; i.e., $u_{01}(t_{8})=u_{03}(t_{8})=\ep\ll1 $. 
Expanding the equations for forward iteration in $(u_{61}, v_{61})$, we see that in the coordinates of the first chart
\be
	\left\{\begin{array}{rcl} \wb{u}_{61}(t_{8}) & = & -t_{8}v_{01}(t_{8})+{\mathcal O}(\ep),\\ 
					   \wb{v}_{61}(t_{8}) & \sim & u_{01}(t_{8})={\mathcal O}(\ep) \end{array} \right. 
\ee
and in the coordinates of the second chart
\be
	\left\{\begin{array}{rcl} \wb{u}_{61}(t_{8}) & = & -\dfrac{t_{8}}{v_{03}(t_{8})}+{\mathcal O}(\ep)\\ 
					   \wb{v}_{61}(t_{8}) & \sim & u_{03}(t_{8})={\mathcal O}(\ep). \end{array} \right. 
\ee
The image lies near the line $v_{61}=0$, or rather the component $D_{6}$.

\section{Mappings of remaining exceptional lines}\label{appendix_elines} 
In this appendix, we analyse the dynamics starting with initial values near the exceptional lines $E_{1},E_{2},E_{7},E_{8}$. For simplicity and conciseness, we focus on the local results of iterating a neighbourhood near each exceptional line and, to do so, we assume that initial values in a  neighbourhood of $E_i$ are analytic functions of $t$ close to a point $t_i$,  for each $i=1,2,7, 8$. Being analytic away from singularities, the birational map \eqref{qp1} maps a disk near an exceptional line to another disk of non-zero size. 

\subsection{Behaviour near $E_{1}$}
The exceptional line $E_{1}$ arises under blow-up of the point $b_{1}$ in the region where $u,v$ are finite; it is the line $v_{11}=0$, or equivalently, $u_{12}=0$. Expanding the equations \eqref{qp1} for forward and backward iteration, we find in the coordinates of the first chart:
\be
	\left\{\begin{array}{rcl} \wb{u}_{01}(t_{1}) & \sim & t_1\,\dfrac{qu_{11}(t_{1}) - t_{1}^3}{qu_{11}(t_{1})^2},\\ 
					   \wb{v}_{01}(t_{1}) & \sim & \dfrac{u_{11}(t_{1})}{t_{1}^2}, \end{array} \right. \quad\quad\quad
	\left\{\begin{array}{rcl} \ub{u}_{71}(t_{1}) & \sim & -qt_{1}(1+qu_{11}(t_{1})),\\ 
					   \ub{v}_{71}(t_{1}) & \sim & -\,\dfrac{v_{11}(t_{1})}{q}, \end{array} \right. 
\ee 
and in the coordinates of the second chart:
\be
	\left\{\begin{array}{rcl} \wb{u}_{01}(t_{1}) & \sim & \dfrac{t_{1}v_{12}(t_{1})(q - v_{12}(t_{1})t_{1}^3)}{q},\\ 
					   \wb{v}_{01}(t_{1}) & \sim & \dfrac{1}{t_{1}^2v_{12}(t_{1})},\end{array} \right. \quad
	\left\{\begin{array}{rcl} \ub{u}_{71}(t_{1}) & \sim & -\dfrac{q(q+v_{12}(t_{1}))}{t_{1} v_{12}(t_{1})},\\ 
					   \ub{v}_{71}(t_{1}) & \sim & -\, \dfrac{v_{12}(t_{1})u_{12}(t_{1})}{q}. \end{array} \right. 
\ee 
The image under the forward map is a curve in regular space.
Under backward iteration, we are mapped to near the line $v_{71}=0$, i.e., to $E_{7}$.


\subsection{Behaviour near $E_{2}$}
The exceptional line $E_{2}$ is the line $v_{21}=0$, $u_{22}=0$. 
Suppose that at a time $t_{2}$ we are near this exceptional line; i.e., $v_{21}(t_{2})$, $u_{22}(t_{2})$ are close to zero.
Expanding the equations for $(u_{81}, v_{81})$ expressed in terms of the coordinates $(u_{21}, v_{21})$, we find
\be
\left\{\begin{array}{rcl}
		 \wb{u}_{81}(t_{2}) & \sim & -\dfrac{(q u_{21}(t_{2})+1)}{t_{2}u_{21}(t_{2})},\\
 		 \wb{v}_{81}(t_{2}) & \sim & u_{21}(t_{2})v_{21}(t_{2}), \end{array}\right. \quad\quad
\left\{\begin{array}{rcl}
		 \ub{u}_{01}(t_{2}) & \sim & \dfrac{t_{2}^2}{q^2u_{21}(t_{2})},\\
 		 \ub{v}_{01}(t_{2}) & \sim & q t_{2} u_{21}(t_{2})(1 - q^4 t_{2}^3 u_{21}(t_{2}), \end{array}\right. 
\ee
while in the coordinates of the second chart, we obtain
\be
\left\{\begin{array}{rcl}
		 \wb{u}_{81}(t_{2}) & \sim & -\dfrac{q+v_{22}(t_{2})}{t_{2}},\\
 		 \wb{v}_{81}(t_{2}) & \sim & u_{22}(t_{2}), \end{array}\right. \quad\quad\quad
\left\{\begin{array}{rcl}
		 \ub{u}_{01}(t_{2}) & \sim & \dfrac{v_{22}(t_{2})}{q^2 t_{2}^2},\\
 		 \ub{v}_{01}(t_{2}) & \sim & \dfrac{q t_{2}(v_{22}(t_{2})-q^4t_{2}^3)}{v_{22}(t_{2})^2}. \end{array}\right. 
\ee
We see that under forward iteration the image lies near the line $v_{81}=0$, i.e., near $E_{8}$.
Under backward iteration, this is a curve in regular space.


\subsection{Behaviour near $E_{7}$} 
$E_{7}$ is the line $v_{71}=0$ or equivalently $u_{72}=0$. Suppose that at a time $t_{7}$ we are near this exceptional line; i.e., $v_{71}(t_{7})$, $u_{72}(t_{7})$ are close to zero. 
Expanding the equations for $(u_{11}, v_{11})$ in coordinates $(u_{71}, v_{71})$, we find:
\be
	\left\{\begin{array}{rcl} \wb{u}_{11}(t_{7}) & \sim & -\dfrac{q^3 + u_{71}(t_{7})t_{7}}{q^4},\\ 
					   \wb{v}_{11}(t_{7}) & \sim & -q v_{71}(t_{7}), \end{array} \right. \quad\quad\quad
\ee
while the backward map for $(u, v)$ in $(u_{71}, v_{71})$ gives
\be
	\left\{\begin{array}{rcl} \ub{u}_{01}(t_{7}) & \sim & q^2 t_{7},\\ 
					   \ub{v}_{01}(t_{7}) & \sim & -\dfrac{(1+t_{7} u_{71}(t_{7}))}{q^3 t_{7}^2}, \end{array} \right. \quad\quad\quad
\ee
Now consider these maps in in coordinates $(u_{72}, v_{72})$:
\be
	\left\{\begin{array}{rcl} \wb{u}_{11}(t_{7}) & \sim & -\dfrac{q^3v_{72}(t_{7}) + t_{7}}{q^4v_{72}(t_{7})},\\ 
					   \wb{v}_{11}(t_{7}) & \sim & -\,q v_{72}(t_{7})u_{72}(t_{7}), \end{array} \right. \quad\quad\quad
\ee
\be
	\left\{\begin{array}{rcl} \ub{u}_{01}(t_{7}) & \sim & q^2 t_{7},\\ 
					   \ub{v}_{01}(t_{7}) & \sim & -\,\dfrac{(t_{7}+v_{72}(t_{7}))}{q^3 t_{7}^2 v_{72}(t_{7})}. \end{array} \right. \quad\quad\quad
\ee
Under forward iteration the image lies near the line $v_{11}=0$, or rather $E_{1}$ shifted forward.
Under backward iteration, we are mapped to near the line $u_{01}=q^2t$.


\subsection{Behaviour near $E_{8}$} 
The exceptional line $E_{8}$ is the line $v_{81}=0$, $u_{82}=0$. 
Suppose that at a time $t_{8}$, we are near this exceptional line; i.e., $v_{81}(t_{8})$, $u_{82}(t_{8})$ are close to zero.
Expanding the forward and backward maps in chart 1 for small $v_{81}(t_{8})$, $u_{82}(t_{8})$, we find in coordinates $(u_{81}, v_{81})$
\be
	\left\{\begin{array}{rcl} \wb{u}_{01}(t_{8}) & \sim & -\dfrac{(q^2+t_8\,u_{81}(t_{8}))}{q t_8^2},\\ 
					   \wb{v}_{01}(t_{8}) & \sim & \dfrac{t_8}{q}, \end{array} \right. \quad
	\left\{\begin{array}{rcl} \ub{u}_{21}(t_{8}) & \sim & -\dfrac{1}{q\,(1+q t_{8} u_{81})},\\ 
					   \ub{v}_{21}(t_{8}) & \sim &-\, q\,(1+q t_{8} u_{81})v_{81}, \end{array} \right. 
\ee
while in the coordinates of the second chart
\be
	\left\{\begin{array}{rcl} \wb{u}_{01}(t_{8}) & \sim & -\dfrac{q^2 v_{82}(t_{8})+1)}{q t_8^2 v_{82}(t_{8})},\\ 
					   \wb{v}_{01}(t_{8}) & \sim & \dfrac{t_8}{q}, \end{array} \right. \quad
	\left\{\begin{array}{rcl} \ub{u}_{21}(t_{8}) & \sim & -\dfrac{v_{82}(t_{8})}{q(q \,t_8+v_{82})},\\ 
					   \ub{v}_{21}(t_{8}) & \sim & q\,(q \,t_8+v_{82}) u_{82}. \end{array} \right. 
\ee
We see that under forward iteration this is near the line $v_{01}=t/q$.
Under backward iteration, we are mapped to near the line $v_{21}=0$, or rather $E_{2}$ shifted backward.

\section*{Acknowledgements} The authors would like to thank H. Dullin, A. Dzhamay, C. Lustri and T. Takenawa for informative discussions. The research reported in this paper was supported by Australian Laureate Fellowship Grant \#FL120100094 from the Australian Research Council.

\bibliographystyle{amsplain}

\end{document}